\documentclass[journal]{IEEEtran}
\usepackage{amssymb,amsmath,amsfonts,amssymb,graphicx}
\usepackage[justification=centering]{caption}
\ifCLASSINFOpdf
\else
\fi
\begin{document}
%
\title{Assured Data Deletion with Fine-grained Access Control for Fog-based Industrial Applications}
%


\author{Yong~Yu, 
        Liang~Xue, 
				Yannan~Li$^{\ast}$\thanks{$^\ast$ Corresponding Author}, 
				Xiaojiang~Du$^{\ast}$, 
				Mohsen~Guizani, 
        and Bo~Yang
				
				
\IEEEcompsocitemizethanks{\IEEEcompsocthanksitem Yong Yu, Liang Xue, Yannan Li and Bo Yang are with School of Computer Science, Shaanxi Normal University, Xi'an, 710062, China. (E-mail: \{yuyong, byang\}@snnu.edu.cn, \{isliangxue, liyannan2016\}@163.com)
\IEEEcompsocthanksitem
Yong Yu and Bo Yang are with Shanghai KOAL Software Co., Ltd, Shanghai, 200436, China.
\IEEEcompsocthanksitem
Xiaojiang Du is with Dept. of Computer and Information Sciences, Temple University, USA. (Email: dxj@ieee.org)
\IEEEcompsocthanksitem
Mohsen~Guizani is with Department of Electrical and Computer Engineering, University of Idaho, USA. (Email: mguizani@ieee.org)
}

}

\maketitle
\maketitle

\begin{abstract}
The advances of cloud computing, fog computing and Internet of Things (IoT) make the industries more prosperous than ever. A wide range of industrial systems such as transportation systems and manufacturing systems have been developed by integrating cloud computing, fog computing and IoT successfully. Security and privacy issues are a major concern that hinders the wide adoptions of these novel techniques. In this paper, we focus on assured data deletion, an issue which is important but received less attention in academia and industry. We firstly propose a framework to integrate the cloud, the fog and the things together to manage the stored data from industries or individuals. We then focus on secure data deletion in this framework by proposing an assured data deletion scheme which fulfills fine-grained access control over sensitive data and verifiable data deletion. Only the data owners and the fog devices are involved when deleting a data key and validating the data deletion, which makes the protocol practical due to the features of low latency and real-time interaction of fog computing. The proposed protocol takes advantage of attribute-based encryption and is provably secure under the standard model. The theoretical analysis shows the good performance and functionality requirements while the implementation results demonstrate the feasibility of our proposal.

%
\end{abstract}


%
\IEEEpeerreviewmaketitle

\section{Introduction}
%
%
%
%
\IEEEPARstart{C}loud computing \cite{Mell2009}, a type of novel computing paradigm which provides shared processing resources and data to computational devices on demand, has made our lives more convenient and easier. For example, the cloud can turn a resource-constrained mobile device into a supercomputer, which means that we can get super computational power to analyse virtually any type of information wherever we are, from the cloud. As a result, people can benefit a lot by employing cloud computing services. We take software as a service as an instance. Currently, people are enjoying the convenient services provided by Google map, Facebook, Youtube and so on. Cloud computing has a number of benefits such as savings in the cost of upgrades, easy scalability, flexibility in data access, regular backups and disaster recovery. However, there exist two drawbacks namely security and centralization in cloud computing \cite{cloud1}.

Cloud computing clearly creates effective economies of scale by using simple centralized architectures. However, traditional cloud computing centralization \cite{TDSC,tifs1,tifs2}, including centralized data storage and data processing, may lead to some limitations. The first limitation is the loss of privacy \cite{Ducloud} due to outsourcing data to a centralized server and the second one is centralization which may cause unreliable latency, lack of mobility support and location-awareness. These inherent problems of cloud computing might hamper human-centered designs, which allow blurring the boundaries between human and machines to emerge more interesting applications.

Amounts of data are produced at the edge of the network, thus, it is more efficient to do computations on the data at the edge of the network. Several emerging techniques such as cloudlet and fog computing \cite{Wen} have been developed since cloud computing is not always efficient enough for data processing when data is generated at the edge of the network. Fog computing has emerged as a promising technology that can push the frontier of computing applications, data processing, and services out of the centralized data centers and down to the logical periphery of the network. This is extremely beneficial when considering the decreased pressure on network bandwidth and communication latency.

Internet has provided so much convenience to our modern society. People in today's generation are relying on the internet for many different tasks, e.g., work, entertainment, shopping. It has been a common trend that electronic devices are equipped with Internet connection. Internet of things (IoT) \cite{IoT} is a collection of ``things'' embedded with electronics, software, sensors, actuators, and connected via the Internet to collect and exchange data with each other.
The applications and services running on IoT devices \cite{iot1}\cite{iot2} are connected to the Internet and may need to send their collected data to powerful servers for processing or analytics. If these aggregated data are in huge volume and the processing task is time-sensitive, the traditional centralized processing would cause a heavy burden on network bandwidth and also may fail the real-time requirement considering the distance between the central data center and the requestor. Fog computing, instead, shifts the workloads away from the centralized data centers and clouds (i.e., ``core'' nodes) to the decentralized fog nodes deployed geographically near the IoT devices that requested for resources.


Fog computing is a highly virtualized platform that provides storage, computing and networking services between the end devices and cloud centers \cite{fog1}. The advantages of fog computing such as location awareness, low latency, support for mobility and predominance of wireless access make it a suitable platform for applications in many domains, such as IoT, software defined networks, wireless sensors and actuators networks. In the framework of fog computing \cite{Wen}, each smart thing is attached to one of fog devices. Fog devices can be interconnected and each of them is linked to the cloud. There is a fruitful interplay between the cloud and fog, especially when it comes to data management and analytics.

Fog computing offers a promising opportunity to build powerful industrial systems by taking advantage of the growing ubiquity of smart things of IoT \cite{Xu}\cite{fog2}. For example, IoT has been adopted in the healthcare service industry to improve healthcare service quality.
In healthcare scenario, medical sensors carried by patients monitor the parameters such as blood pressure, breathing activity that can be processed and provide guidance to the healthcare staff. Data privacy here is major concern that hinders the general adoption of fog computing. For example, the data should be accessible only to authorized users for the reason that obtaining inaccurate and wrong medical data may lead to wrong treatments and it will cause privacy violation if everyone can access the data. Broadcast encryption \cite{broad1}, \cite{broad2}, \cite{broad3}, \cite{broad4} is a choice to achieve the goal that an encrypted data can be decrypted by a number of different authorized users, but it requires an encryptor to have the prior knowledge of every prospective receiver as well as the authorization information associated with each receiver. Attribute-based encryption \cite{KPABE}, \cite{CPABE} is a highly promising cryptographic primitive to achieve fine-grained access control \cite{access1,access2,access3,access4}, by which only the users whose attributes satisfy the access structure embedded in the ciphertext can decrypt the ciphertext.

Besides fine-grained access control, data security is another concern in fog computing since a user will lose control over their outsourced data. Prior research mainly focuses on the existence of the data, i.e., data integrity, which has been well resolved. We investigate a complementary problem, that is, when a patient wants to delete his sensitive medical data from an outsourced storage, how can he guarantee that the deleted data will never resurface if he does not delete the data by himself? This is a well-recognized issue but received less attention than data integrity. A privacy-focused service may wish to promptly and securely delete the data once they have served their purpose. Network services may also need secure data deletion simply to comply with regulations regarding their users' sensitive data. In 2014, the European Union's  \emph{right to be forgotten} forces companies to store users' data in a manner that supports secure data deletion upon request. Meanwhile, California's legislation enforces similar requirements. Assured data deletion is a natural and necessary requirement in Fog computing especially for the confidential data. However, secure data deletion is challenging since it is much harder to prove \emph{non-existence} of outsourced data than to prove the \emph{existence} of the data.
The goal of assured data deletion \cite{Readon} is deleting data from a storage medium, so as to make the data irrecoverable and prevent the adversary from gaining access to them.

In the physical world, secure deletion can be realized easily: confidential documents are shredded, sensitive information of a file is selectively redacted. In the digital world, however, assured data deletion seems simple but there are many tough issues under the surface. Medium's physical destruction such as incineration or pulverization is an accepted approach if the storage medium does not need to be reused. However, secure deletion is not one-off event which deletes the all stored data. Instead, it is common that secure deletion needs to be implemented at a fine granularity and the system needs to continue functioning normally. In fog computing, it needs the trust to the platform since there is no technical proof, and the infrastructure being out of users' control makes them difficult to verify the deletion of data. Now, file systems may implement deletion simply by unlinking the files due to it can be done by changing even only one bit. But this will result in the full contents remain available. Deleted data can be regained by means of data recovery technique, thus cannot achieve the assured data deletion.

The cryptography-based approaches are encrypting data before storage, thus, the data deletion problem is converted to secure erasure of the related secret key. Perman \cite{Perlman} proposed the first assured data deletion file system where each file is created with an expiration time. In \cite{Perlman}, the file is securely deleted by removing the ephemeral keys corresponding to the expiration time. However, it needs to determine the expiration time of the key in advance. Moreover, the deletion of data totally depends on the trusted Ephemerizer servers, which may lead to single point failures. Geambasu et al. \cite{Geambasu} presented the Vanish, which is designed for the secure deletion of communication over the Internet. In Vanish system, each message is encrypted with a key, and the key is divided into shares and stored in the Distributes Hash Table (DHT) network. Nodes in the DHT remove the key shares that reside in their caches for a fixed time, which leads to the message inaccessible. However, in Vanish system, the deletion is constrained by the node's update cycle, and Vanish cannot resist the Hopping attack and Sniffing attack. The Hopping adversary can change the port number from time to time to join the network to replicate the associated key components (usually 16-51 bytes). By sniffing, the components of a key in transit can be easily captured by the adversary. In 2012, Tang et al. \cite{Tang} generalized the time-based deletion to policy-based deletion. In their scheme, each file is associated with a boolean combination of atomic policies. A file is encrypted with a data key, and the data key is further encrypted with the control keys which are associated with policies. By revoking the policy, the corresponding control key is deleted from the key manager \cite{key1}, who is responsible for key management \cite{Dukey1,Dukey2, Dukey3}. The flaw of the FADE is that the key managers need to perform complex decryption operations for each message, so it is not suitable for large, dynamic user scenarios. Perito and TSudik designed an assured deletion scheme \cite{Perito} for embedded devices with limited memory. It makes use of secure code update which securely erase all data on the device and then download the new code, but it fails to achieve fine-grained, flexible deletion. Mo \cite{Mo} proposed a fine-grained assured deletion scheme by using key modulation function. In the scheme, all data keys are derived from master key. When a data key $k$ needs to be deleted, the master key is changed so that the $k$ is unrecoverable, while other keys are attainable from the master key via running the modulator adjustment algorithm. Readon et al. \cite{ReardonJ} used B tree to organize, access and securely delete data. Both the data and B-Tree are stored on the persistent storage medium in their scheme, and by introducing a shadowing graph mutation, they prove that it achieves secure deletion of data. But it needs to create the arborescent structure for the data.

\textbf{Our Contributions.} To meet the requirements of fine-grained access control and secure data deletion, in this paper, we propose a ciphertext-policy based assured deletion (CPAD) scheme. Smart objects are usually deployed for specific applications, which can be utilized to describe data accessibility by encrypting the data with attributes owned by potential users. The access structure in the ciphertext can be sophisticated logic expressions on the attributes. In our protocol, considering that the smart objects are usually resource-limited and expensive cryptographic primitives cannot be directly employed, thus data are encrypted via symmetric encryption algorithm such as AES. We call the key used in the symmetric encryption the data key. Data key is then encrypted with an access structure. The ciphertext of the data key and encrypted data are uploaded to the closest fog device. The fog device then stores the ciphertext of data key but uploads the ciphertext of data to the cloud. We can realize the assured data deletion by attribute revocation such that no user satisfies the access structure in the ciphertext. Moreover, the fog device can respond a proof by which the smart object can verify that the data are deleted indeed. In summary, the contributions of this paper are three-folds.

\begin{itemize}
  \item The proposed protocol achieves fine-grained access control and assured deletion for data generated by smart objects. Users are assigned a series of attributes. A ciphertext is associated with an access structure, which supports any linear secret sharing schemes. A ciphertext can be decrypted only if the user's attributes meet the access structure in the ciphertext. Assured data deletion is achieved by revoking an attribute in the ciphertext. The fog device can return a proof, and the smart object is convinced that the data are deleted if the received proof is valid.

  \item The proposed protocol is provably secure under the Decisional $q$-Parallel BDHE assumption. It is resistant against unauthorized users' collusion attack, that is, the collusion of unauthorized users cannot lead to the disclosure of data. Moreover, we remove the centralized trusted service, i.e., key manager, in our protocol.


  \item We empirically evaluate the performance overhead of the proposed protocol, which demonstrates CPAD is efficient in terms of computation and communication overhead. The implementation of the protocol shows the practicability of CPAD in the real-world applications.
\end{itemize}

\textbf{Organization:} The remainder of the paper is organized as follows. In section \ref{sec2}, we review some preliminaries used in our construction. In section \ref{sec3}, we describe the system model, adversary model and security requirements of our assured data deletion protocol. We present our concrete construction in section \ref{sec4}, followed by the security analysis of our scheme in section \ref{sec5}. Performance evaluation and implementation results are given in section \ref{sec6}. We conclude this paper in section \ref{sec7}.

\section{Preliminaries}\label{sec2}

In this section, we review some fundamental backgrounds used in the protocol, including bilinear pairings, Decisional Parallel $BDHE$ assumption, access structure, linear secret sharing schemes and attribute-based encryption.

\subsection{Bilinear Pairing and Complexity Assumptions}
\noindent{\textbf{Definition 1. (Bilinear Pairing)}. A bilinear pairing \cite{IDWP} is a map $e: G \times G\rightarrow G_T$, where $G$, $G_T$ are two multiplicative cyclic bilinear groups of prime order $p$. The map has the following properties.}

\textbf{Bilinearity.}  For all $h_1,h_2 \in G$, and $a,b \in \mathbb{Z}_p$, $e(h_1^a, h_2^b)=e(h_1,h_2)^{ab}$.

\textbf{Non-Degeneracy.}  $e(g,g) \notin 1$,where $g$ is a generator of $G$.

\textbf{Efficient Computation.} There exists an efficiently computable algorithm to compute $e(h_1,h_2)$ for all $h_1,h_2 \in G$.

\noindent{\textbf{Definition 2. (Decisional $q$-Parallel Bilinear Diffie-Hellman Exponent assumption \cite{Waters})}. Choose a group $G$ of prime order $p$. Let $g$ be the generator of $G$ and $a,s,b_1,\dots b_q$ be chosen randomly from $\mathbb{Z}_p$. If an adversary is given $\vec{y}$ $=$
$$g,g^s,g^a,\cdots,g^{a^q},,g^{a^{q+2}},\cdots,g^{a^{2q}}$$
$$\forall{1\leq j \leq q},~g^{s\cdot b_j},g^{a/b_j},\cdots,g^{a^q/b_j},,g^{a^{q+2}/b_j},\cdots,g^{a^{2q}/b_j}$$
$$\forall{1\leq k,j\leq q, k\neq j,}~g^{a\cdot s\cdot b_k/b_j},\cdots,g^{a^q\cdot s\cdot b_k/b_j}$$

it is hard to distinguish the $e(g,g)^{a^{q+1}s}$ from a random element $R$ in $G_T$. We say an algorithm $\mathcal{B}$ that outputs $z\in \{0,1\}$ has advantage $\epsilon$ in solving Decisional $q$-$Parallel$ $BDHE$ problem if
$$|Pr[\mathcal{B}(\vec{y},e(g,g)^{a^{q+1}s})=0]-Pr[\mathcal{B}(\vec{y},R)=0]|\geq \epsilon$$

The Decisional $q$-$Parallel$ $BDHE$ assumption holds if no adversary has a non-negligible advantage in solving the Decisional $q$-$Parallel$ $BDHE$ problem. The proof that the assumption generally holds is in \cite{Waters}}.

\subsection{Access Structures}
\noindent{\textbf{Definition 3. (Access Structure \cite{LSSS})}. Let $\{P_1,P_2,\dots P_n\}$ be a set of parties. A collection $\mathbb{A}\subseteq 2^{\{P_1,P_2,\dots P_n\}}$ is monotone if $\forall~B,C$ : if $B \in \mathbb{A}$ and $B\subseteq C$, then $C \in \mathbb{A}$. An access structure (respectively, monotone access structure) is a collection (respectively, monotone collection) $\mathbb{A}$ of non-empty subsets of $\{P_1,P_2,\dots P_n\}$, i.e. , $\mathbb{A}\subseteq 2^{\{P_1,P_2,\dots P_n\}}\backslash \{\varnothing\}$. The sets in $\mathbb{A}$ are called the authorized sets, and the sets not in $\mathbb{A}$ are called the unauthorized sets.

In our protocol, the parties are identified by the attributes. Thus the access structure $\mathbb{A}$ contains the authorized sets of attributes.}

\subsection{Linear Secret Sharing Schemes}
\noindent{\textbf{Definition 4. (Linear Secret Sharing Schemes (LSSS)\cite{LSSS})}. We recall the definitions of LSSS given in \cite{LSSS} as follows. A secret sharing scheme $\Pi$ over a set of parties $P$ is linear if

\begin{itemize}
  \item\sl{The shares for each party form a vector over $\mathbb{Z}_p$.}
  \item\sl{There exists a matrix $M$ with $l$ rows and $n$ columns called the share-generating matrix for $\Pi$. For $i=$1$,\dots, l$, $\rho(i)$ is a function, which defines the party labeling row $i$. For the column vector $\vec{v}=(s,r_2,\dots,r_n)$, where $s\in \mathbb{Z}_p$ is the secret to be shared, and $r_2,\dots,r_n \in \mathbb{Z}_p$ are chosen randomly, $M\vec{v}$ is the vector of $l$ shares of the secret s according to $\Pi$. The share $M\vec{v}_i$ belongs to party $\rho(i)$.}
\end{itemize}

Linear secret sharing scheme enjoys the \emph{linear reconstruction} property: Suppose that $\Pi$ is an LSSS for the access structure $\mathbb{A}$ and $S \in \mathbb{A}$ is any authorized set. Let $I\subset \{1,2,\dots l\}$ be the set that $I=\{i:\rho(i)\in S\}$. Then, if $\{\lambda_i\}_{i\in I}$ are valid shares of secret $s$ according to $\Pi$, there exist constants $\{\omega_i \in \mathbb{Z}_p\}_{i\in I}$ such that $\sum_{i\in I}\omega_i\lambda_i=s$. And these constants $\{\omega_i \in \mathbb{Z}_p\}_{i\in I}$ can be found out in time polynomial in the size of share-generating matrix $M$.

Note: Using standard techniques \cite{LSSS}, one can convert any monotonic boolean formula (including access formulas in terms of binary trees) into an LSSS representation. In our protocol, we use the vector $(1,0,0\dots,0)$ as the target vector for linear secret sharing schemes. For any satisfying set of rows $I$ of $M$, the target vector is in the span of $I$. For any unauthorized set of rows $I$ of $M$, the target vector is not in the span of $I$. Moreover, there exists a vector $\vec{\omega}$ such that $\vec{\omega}\cdot (1,0,0\dots,0)=-1$ and $\omega\cdot \vec{M}_i=0$ for all $i\in I$, where $\vec{M}_i$ is the $i$-th row of the matrix $M$.

\subsection{Attribute-based encryption}
The first attribute-based encryption scheme \cite{Fuzzy05} was proposed by Sahai and Waters. In attribute-based encryption schemes, attributes have been exploited to generate Public/Private key pairs for users, and have been used as an access policy to control users' access. Depending on the access policy, attribute-based encryption schemes can be roughly categorized as key-policy attribute-based encryption (KP-ABE) and ciphertext policy attribute-based encryption (CP-ABE) schemes.

The CP-ABE builds the access structure in the ciphertext which can choose the corresponding user's private key to decipher encrypted data. It improves the disadvantages of KP-ABE that the data owner cannot choose who can decrypt the ciphertext. Moreover, a users' private key corresponding to a set of attributes, so the user can decrypt the ciphertext whith the attributes, which satisfies the access structure. Thus, CP-ABE is very close to traditional access control and can be applied in a number of real-world applications.

In 2007, Bethencourt et al. \cite{CPABE} proposed the first ciphertext-policy attribute-based encryption scheme, which consists of four algorithms described below.

\emph{Setup($\lambda$)}: This algorithm takes a security parameter $\lambda$ as input. It outputs the public parameters $PP$ and a master key $MK$.

\emph{Encrypt($PK,M,\mathbb{A}$)}: This algorithm takes as input the public parameter $PK$, a message $M$ and an access structure $\mathbb{A}$. The message is encrypted with the access structure $\mathbb{A}$. This algorithm outputs the ciphertext $CT$ of $M$.

\emph{KeyGen($MK,S$)}: This algorithm takes as input the master key $MK$ and a set of attributes $S$ and outputs the private key $SK$.

\emph{Decrypt($CT,PK,SK$)}: This algorithm takes as input the ciphertext $CT$ which implicitly contains an access structure $\mathbb{A}$, public parameters $PP$ and a private key $SK$, which is a private key for a set of attributes $S$. If the set of attributes satisfy the access structure $\mathbb{A}$, then this algorithm decrypts the ciphertext and outputs the message $M$.

\section{Models and Requirements}\label{sec3}
In this section, we describe the system model, security requirements and security model in fog based assured data deletion.

\subsection{System Model}
We consider the scenario where there are many smart objects generating or collecting data. Each smart thing is connected to one of fog devices by a wireless access. Fog devices could be interconnected and each of them is linked to the cloud as shown in Fig. 1. There is an Attribute Authority (AA), who is responsible for the system initialization and attributes management in the system. AA publishes the system parameters which are needed in encryption and assigns private keys to the users. The cloud, which has powerful computing and storage resources, provides data storage, access control and data processing services for users. The fog devices allow applications to run as close as possible to sensed, actionable and massive data. It provides compute, storage and networking services between end devices and traditional cloud servers. We assume fog nodes are honest but curious, which means it honestly executes the tasks assigned by legitimate parties in the system. However, it would like to learn information of the data in the cloud as much as possible. Smart objects, such as sensors or embedded devices in IoT, generate data and upload encrypted data to the fog devices and the cloud. Smart objects cannot afford expensive computational operations due to limited computing and storage capacity. Users, whose attributes satisfy the access structure in the ciphertext, can recover the data.

\begin{figure}[ht]
 \centering
 \includegraphics[width=0.5\textwidth]{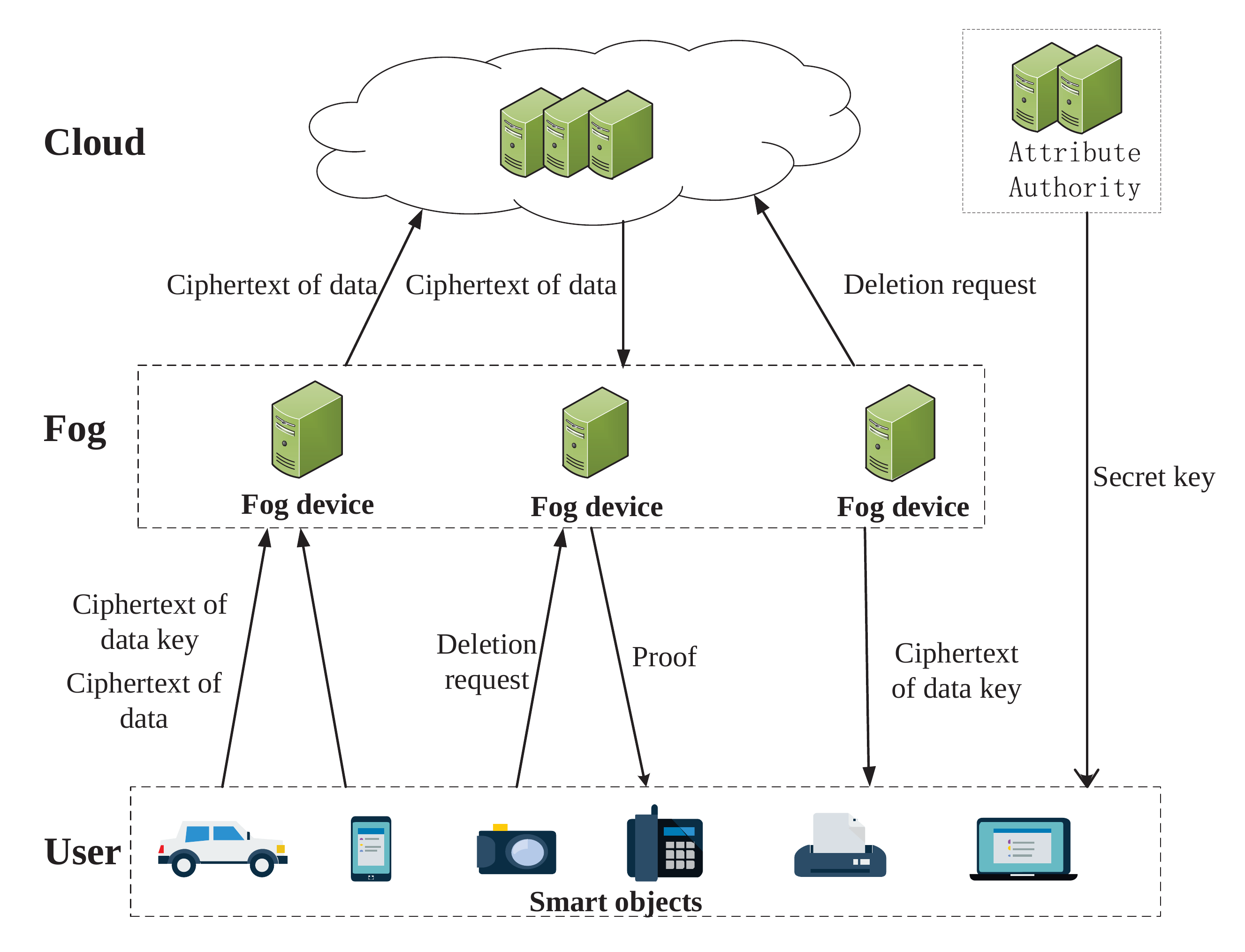}
 \caption{ System model}
\end{figure}


\subsection{Security Requirements}

 We consider adversaries whose main goal is to obtain the data which they are not authorized to access. The attackers can eavesdrop all the communications in the system, and unauthorized users may collude to compromise the encrytped data. After the data are deleted, adversaries try to recover the deleted data. The adversaries here include both authorized users and unauthorized users. We assume fog devices are honest, hence we do not consider attacks in which users collude with fog devices.

For secure data storage and access, some common security properties such as data confidentiality and integrity, should be guaranteed. Here, we focus on two specific goals that CPAD seeks to achieve: secure fine-grained access control and assured deletion of data.

Secure fine-grained access control: As mentioned previously, fine-grained data access control is always desirable in many applications. The proper scheme should be able to precisely specify the capability of different kinds of users to access the data at different security levels. A user can decrypt the ciphertext only when the attributes of the user satisfy the access structure. Unauthorized users can not get access to the sensitive data even they collude with each other.

Assured data deletion: The data should be permanently inaccessible to users after data deletion. An adversary cannot get any knowledge of the data even it can access the data previously. The fog devices should return a deletion proof with which the smart object can validate that the data have been indeed deleted.

\subsection{Security model}
In the security model of assured deletion, the adversary chooses an access structure $\mathbb{A^*}$ at the beginning of the game, such that the data key encrypted under the access structure $\mathbb{A^*}$ is deleted in the end. The adversary can also query any secret keys corresponding to the set of attribute $S$, such that $S$ does not satisfy $\mathbb{A^*}$.

The security of the ciphertext-policy based assured data deletion scheme is defined using the following game in which an adversary and a challenger are involved.

\noindent{\textbf{Init}: The adversary outputs an access structure $\mathbb{A^*}$ in which the dummy attribute is revoked, meaning the data key encrypted under the access structure $\mathbb{A^*}$ is deleted.}

\noindent{\textbf{Setup}: The challenger runs the Setup algorithm and give the adversary public parameters $PK$.}

\noindent{\textbf{Phase 1}: The adversary makes a number of private key queries corresponding to sets of attributes $S_1,\dots,S_{q_1}$. The restriction for each query is that none of the attribute set $S_i$ satisfies $\mathbb{A^*}$. Since the attribute ``dummy'' is revoked, the set of attribute $S_i$ does not contain the attribute ``dummy''.}

\noindent{\textbf{Challenge}: The adversary outputs two equal length messages $M_0, M_1$ and sends them to the challenger. The challenger picks a random bit $b\in \{0,1\}$ and encrypts the message $M_b$ under $\mathbb{A^*}$. Then the ciphertext $CT^*$ is forward to the adversary.

\noindent{\textbf{Phase 2}: The adversary issues additional private key queries corresponding to the sets of attributes $S_{q_1+1},\dots,S_q$ with the same restrictions as in Phase 1.}

\noindent{\textbf{Guess}:  The adversary outputs a guess $b' \in \{0,1\}$ and wins if $b=b'$.}

The advantage of the adversary in attacking the game is defined as $Adv(\mathcal{A})=|Pr[b=b']-\frac{1}{2}|$. A ciphertext-policy based assured data deletion scheme is secure if no polynomial time adversary has a non-negligible success advantage in the aforementioned game.

\section{Our Construction}\label{sec4}
In this section, we describe a concrete construction that fulfills the properties of assured data deletion and fine-grained data access. In our protocol, we make use of the ciphertext-policy attribute-based encryption. Attribute authority (AA) initializes the system and publishes system parameters. We denote the universe of all the attributes in the system by a symbol $U$. There is an attribute ``dummy'', which is included in the set of attributes for each user and is an indispensable attribute in the access formula. Each smart object is preloaded the public parameter $PP$ as well as a LSSS access matrix $M$. The data generated by the smart object are encrypted using a symmetric encryption algorithm. The data key is encrypted with the ciphertext-policy attribute-based encryption. The smart object uploads the ciphertext of data and encrypted data key to a fog device, and deletes the local file. The fog device keeps the ciphertext of the data key, and transfers the ciphertext of data to the cloud. When the data generated by the smart objects need to be deleted, the smart object sends a deletion request to the fog device. The smart object and the fog device can agree a secret deletion key via a key exchange protocol. By utilizing the deletion key, the fog device changes the ciphertext related to the attribute ``dummy'', such that users who can access the data previously cannot decrypt the ciphertext. The fog device then sends a deletion request to the cloud for deleting the ciphertext. It is fair to assume the smart object can decrypt the data generated by itself. The smart object can verify the deletion with the deletion key. Our protocol is based on the Waters' ABE construction \cite{Waters}, and AA is not involved in the data deletion process.

\textbf{Setup($1^\lambda$)}. AA chooses two multiplicative cyclic groups $G$ and $G_T$ of prime order $p$ and a bilinear map $e: G\times G \rightarrow G_T$. Let $g$ be a generator of $G$. AA picks random exponents $\alpha,a\in \mathbb{Z}_p$, and chooses a hash function $h:\{0,1\}^*\rightarrow \mathbb{Z}_p$. AA chooses $|U|$ random group elements $h_1,\dots,h_{|U|} \in G$ that are associated with the $|U|$ attributes in the system. The public parameter is published as $PP=\{g,e(g,g)^{\alpha},g^a,h_1,\dots,h_{|U|},h\}$. AA sets $MSK=g^{\alpha}$ as the master secret key.

\textbf{KeyGen($MSK,S$)}: AA takes as input the master secret key $MSK$ and a set of the attributes $S$ of a user, randomly picks $t\in \mathbb{Z}_p$, and computes the private key of the user as follows.
 $$K=g^{\alpha}g^{at},  L=g^t,  \forall x\in S,K_x=h_x^t$$
The private key of the user is $SK=\{K,L,\forall x\in S,K_x\}$.

The smart object generates a signing $public$-$secret$ key pair $\{spk,ssk\}$. The fog device also generates a signing key pair $\{fpk,fsk\}$. The signing algorithm can use short signature where a user or a fog device picks a random $sec \in Z_p^*$, and compute $v=g^{sec}$. It also chooses a hash function $h_1:\{0,1\}^* \rightarrow G \backslash \{1\}$. The $spk$ is $\{v,h_1\}$, and $ssk$ is $sec$.

\textbf{Encrypt($PP,(M,\rho),m,k$)}. The smart object takes as input the public parameter $PP$, data $m$ to be encrypted, an LSSS access structure $AS=(M,\rho)$, and symmetric key $k$. $M$ is an $l*n$ matrix, and $M_i$ denotes the vector corresponding to the $i$-th row of $M$. The function $\rho$ maps each row of $M$ to different attributes. The smart object picks a file name $fname \in \mathbb{Z}_p$ for the data randomly and the data are encrypted with a symmetric encryption such as $AES$. $\{D\}_k$ represents the ciphertext in which $k$ is the symmetric data key. To encrypt the data key, the smart object chooses random values $r_1,\cdots,r_l\in \mathbb{Z}_p$ and a random vector $\vec{v}=(s,y_2,\dots,y_n) \in \mathbb{Z}_p^n$, which is for generating the shares of the encryption exponent $s$. For $i=1$ to $l$, it calculates $\lambda_i=\vec{v}\cdot M_i$. The ciphertext of $k$ is $CT=\{\overline{C},C',\{C_i,D_i\}_{1\leq i\leq l}\}$, where
$$\overline{C}=ke(g,g)^{\alpha s}, C'=g^s$$
$$C_i=g^{a\lambda_i}h_{\rho(i)}^{-r_i}, D_i=g^{r_i}$$

 Then the smart object computes and preserves $\tau=h(fname||k)$, and uploads $\{fname,spk,CT,\{D\}_k,(M,\rho)\}$ to a fog node. Upon receiving the data, the fog node uploads $\{fname,spk,\{D\}_k\}$ to the cloud.

\textbf{Decrypt($CT,\{D\}_k,SK$)}: A user first obtains the ciphertext $CT$ from the fog device and $\{D\}_k$ from the cloud. Suppose the set of attributes $S$ satisfies the access structure $AS$ in the ciphertext, and let $I\subset\{1,2,\dots,l\}$, $I=\{i:\rho(i)\in S\}$, then $\{\omega_i \in \mathbb{Z}_p\}_{i\in I}$ can be calculated where $\{\omega_i\}_{i\in I}$ are constants such that if  $\{\lambda_i\}_{i\in I}$ are valid shares of secret $s$, then $\sum_{i\in I}\omega_i\lambda_i=s$. The decryption process is as follows,
\begin{eqnarray*}
& & \frac{e(C',K)}{\prod_{i\in I}(e(C_i,L)e(D_i,K_{\rho_i}))^{\omega_i}}\\
& = & \frac{e(g,g)^{\alpha s} e(g,g)^{ast}}{\prod_{i\in I}e(g,g)^{ta\lambda_i\omega_i}}\\
& = & e(g,g)^{\alpha s}
\end{eqnarray*}

The user computes $k$ as $k=\frac{\overline{C}}{e(g,g)^{\alpha s}}$, and decrypts the data with $k$.

\textbf{DelRequest($fname,x$)}: $x$ denotes the attribute ``dummy''. When the smart object wants to delete the file named $fname$ on the fog device, it first chooses two random numbers $q,u\in \mathbb{Z}_p$, and computes $\theta=q^u \mod p$. Then the smart object sends the deletion request $DR=\{delete||fname||x||q||\theta,sign_{ssk}(h(delete||fname||x||q||\theta))\}$ to the fog device.

\textbf{ReEncrypt($CT,DR$)}: Upon receiving the deletion request, the fog device verifies the signature in $DR$. If the signature is valid, the fog device transmits the deletion request $DR$ to the cloud. The cloud validates $DR$ again. If the signature in $DR$ is valid, the cloud deletes $\{D\}_k$. The fog device picks a random $v\in \mathbb{Z}_p$, and computes $\eta=q^v \pmod p$, $\gamma=\theta^v \pmod p$. Then the fog device updates the ciphertext $D_i$ in $CT$ for the index $i$ that $\rho_{(i)} = x$, where $x$ denotes the attribute ¡°dummy¡±, as $D'_{x}=D_{x}^{\frac{1}{\gamma}}$ and sends $Resp=\{\eta,sign_{fsk}(\eta)\}$ to the smart object.

\textbf{Verify($Resp,CT'$)}: Upon receiving the response $Resp$ from the fog device, the smart object first verifies the correctness of the signature. If it is valid, the smart object computes $\gamma'=\eta^u$ mod $p$, $K'_{x}=K_{x}^{\gamma'}$, and retrieve the updated ciphertext of $k$ which is $CT'=\{\overline{C},C',\{C_i,D'_i\}_{1\leq i\leq l}\}$ from the fog device. The smart object verifies the deletion as follows.

\begin{center}
for $\rho(i)=x: B_i=(e(C_i,L)e(D'_i,K'_{x}))^{\omega_i}$
\end{center}
\begin{center}
for $\rho(i)\neq x: B_i=(e(C_i,L)e(D'_i,K_{x}))^{\omega_i}$
\end{center}
The smart object computes $A=\prod_{i\in I}B_i=e(g,g)^{ats}$ and $k'=\frac{\overline{C}\dot A}{e(C',K)}$.

Finally, it calculates $\tau'=h(fname||k')$ and compares it with $\tau$. If $\tau'=\tau$ holds, the smart object is convinced that the ciphertext has been changed and the data is inaccessible.

\section{Security Analysis}\label{sec5}
We evaluate the security of the proposed protocol by showing it satisfies all the security requirements described in section \ref{sec3}.

\subsection{Secure Fine-grained Data Access Control}
 Secure fine-grained data access control ensures that only users whose attributes satisfy the access structure in the ciphertext can decrypt the ciphertext correctly. Even the unauthorized users collude, they cannot learn any information from the ciphertext. In our protocol, data are encrypted with a symmetric encryption algorithm. The security of the data can be guaranteed by choosing a secure symmetric encryption algorithm such as AES, which means without the data key, the data is IND-secure. Our encryption of the data key is provably secure under the $q$-$parallel$ $BDHE$ assumption. The proof is similar to that of Waters' scheme \cite{Waters} and so we omit it here. This turns out that the adversary is unable to decrypt the data key unless it owns the intended attributes. Therefore, our protocol can make sure that only authorized users can access the data.

\subsection{Assured data deletion}
Assured data deletion ensures the irrecoverability of the data when the data are deleted. Since data key is encrypted with an access structure and the ciphertext of the data key is re-encrypted, the new ciphertext can not be decrypted correctly by the user's private key, thus, the data is irrecoverable.

When the data are deleted, the security of data key can be proved as follows.

\textbf{Theorem 1}. \emph{Suppose the decisional $q$-$parallel$ $BDHE$ assumption holds in $G$ and $G_T$, there is no polynomial time adversary $\mathcal{A}$ can break the security of our protocol with a non-negligible advantage.}

\textbf{Proof} : The adversary $\mathcal{A}$ claims a challenge matrix $M^*$, and the size of $M^*$ is $l^**n^*$, where $l^*,n^*\leq q$. Suppose the adversary has a non-negligible $\epsilon$ in the selective security game against our protocol, then we can construct a simulator $\mathcal{B}$ that can solve an instance of $q$-$parallel$ $BDHE$ problem with a non-negligible advantage.

$\mathcal{B}$ chooses groups $G$ and $G_T$ with an efficient bilinear map $e$ and a generator $g$. The challenger flips a fair coin $\mu$. If $\mu=0$, the challenger sets the $(\vec{y},T)=(\vec{y},e(g,g)^{a^{q+1}s})$; Otherwise it sets $(\vec{y},T)=(\vec{y},R)$ with a random $R$.

\noindent{\textbf{Init}: $\mathcal{A}$ outputs the challenge access structure $(M^*,\rho^*)$ in which $M^*$ has $n^*$ columns. The data key encrypted under the access structure is deleted in the end.}

\noindent{\textbf{Setup}: $\mathcal{B}$ chooses a random $\alpha' \in \mathbb{Z}_p$, and sets $e(g,g)^{\alpha}=e(g^a,g^{a^q})e(g,g)^{\alpha'}$ which implicitly sets $\alpha=\alpha'+a^{q+1}$.

For each attribute $x$, $\mathcal{B}$ chooses a random value $z_x$. Let $X$ denote the set of indices $i$, such that $\rho^*(i)=x$, which means that the $\rho$ function may be not injective. Then $\mathcal{B}$ programs $h_x$ as follows: If $X=\varnothing$, $h_x=g^{z_x}$; else, $h_x=g^{z_x}\prod_{i\in X}{g^{aM^{*}_{i,1}/b_i} \cdot g^{a^2M^{*}_{i,2}/b_i} \cdots g^{a^{n^*}M^{*}_{i,n^*}/b_i}}$.}

Note that the outputs of the oracle are distributed randomly due to the randomness of $g^{z_x}$.

\noindent{\textbf{Phase 1}: $\mathcal{A}$ adaptively makes requests for the keys. Assume $\mathcal{B}$ is given a private key query for a set of attributes $S$ which does not satisfy $M^*$. The attribute ``dummy'' is not included in the set of attributes $S$ for the reason that the data have been deleted, the ciphertext corresponding to the attribute ``dummy'' is re-encrypted, which leads to the key relevant to the attribute ``dummy'' is invalid. Hence, in the adversary's view, it has no knowledge about the corresponding key.}

$\mathcal{B}$ finds a vector $\vec{\omega}=(\omega_1,\dots,\omega_{n^*})$ such that $\omega_1=-1$, and for all $i$ where $\rho^*(i)\in S, \vec{\omega} \cdot M^{*}_{i}=0$. By the definition of LSSS, such a vector can be found in polynomial time. Then $\mathcal{B}$ chooses a random $r \in \mathbb{Z}_p$ and defines $t$ as $$t=r+\omega_1a^q+\omega_2a^{q-1}+\cdots+ \omega_{n^{*}}a^{q-n^{*}+1}.$$

Then $\mathcal{B}$ sets $L$ as $$L=g^r\prod_{i=1,\cdots,n^*}(g^{a^{q+1-i}})^{\omega_i}=g^t.$$

Since $g^{\alpha}$ is unknown, it seems hard to calculate $K$. However, by the definition of $t$ and $\omega_1=-1$, $g^{at}$ contains a term $g^{-a^{q+1}}$ which can cancel out $g^{\alpha}$ when calculating K. $\mathcal{B}$ computes K as $$K=g^{\alpha'}g^{ar}\prod_{i=2,\cdots,n^*}(g^{a^{q+2-i}})^{\omega_i}.$$

For $\{K_x\}_{x \in S}$, if there is no $i$ such that $\rho^{*}(i)=x$, $\mathcal{B}$ can simply set $K_x=L^{z_x}$. For the $x$ which is used in access structure, the terms of the form $g^{a^{q+1}}/b_i$ are hard to simulate. However, we have $M^*_{i}\cdot \vec{\omega}=0$, all these terms can be canceled.

$\mathcal{B}$ computes $K_x$ as follows.
$$K_x=L^{z_x}\prod_{i\in X}\prod_{j=1,\dots,n^*}\Big( g^{(a^j/b_i)r}\prod_{\substack{K=1,\dots,n^*\\ k\neq j}}(g^{a^{q+1+j-k}/b_i})^{\omega_k}\Big)^{M^*_{i,j}}$$

\noindent{\textbf{Challenge}: $\mathcal{A}$ submits two equal length messages $M_0,M_1$ to $\mathcal{B}$. $\mathcal{B}$ builds the challenge ciphertext as follows.}

$\mathcal{B}$ flips a coin $\beta$. It computes $\overline{C}=M_{\beta}T\cdot e(g^s,g^{\alpha'})$ and $C'=g^s$.

The components $C_i$ contains terms $g^{a^{j}s}$ that $\mathcal{B}$ can't calculate. However, $\mathcal{B}$ can choose the secret splitting, which makes these terms be canceled out. $\mathcal{B}$ chooses random $y'_2,\cdots,y'_{n^*}$ and the vector $\vec{v}=(s,sa+y'_2,sa^2+y'_3,\dots sa^{n-1}+y'_{n^*})\in \mathbb{Z}_p^{n^*}$. Then $\mathcal{B}$ chooses a random key $\eta$ and random values $r'_1,\dots,r'_l$. For the index $i$ that $\rho(i)=x$, where $x$ denotes the attribute ``dummy'', $\mathcal{B}$ sets $D_i=(g^{-r'_i}g^{-sb_i})^{\frac{1}{\eta}}$, otherwise, $D_i=g^{-r'_i}g^{-sb_i}$.
Denote $R_i$ as the set of all $k \neq i,i=1,\dots ,n^*$ such that $\rho^*(i)=\rho^*(k)$, which indicates the set of all other row indices which map to same attributes as row $i$. $C_i$ is generated as $$C_i=h^{r'_i}_{\rho^*(i)}\Big( \prod_{j=2,\dots,n^*}(g^a)^{M^*_{i,j}y'_j}\Big)(g^{b_i\cdot s})^{-z_{\rho^*(i)}}$$ \\ $$\cdot \Big( \prod_{k\in R_i}\prod_{j=1,\dots,n^*}(g^{a^j\cdot s\cdot(b_i/b_k)})^{M^*_{k,j}}\Big)$$
$\mathcal{B}$ sends the adversary $\mathcal{A}$ $CT^*=\{\overline{C},C',\{C_i,D_i\}_{i=1,2,\cdots l^*}\}$ as the challenge ciphertext.

\noindent{\textbf{Phase 2}:  Same as phase 1.}

\noindent{\textbf{Guess}:  Finally, $\mathcal{A}$ outputs a guess $\beta'$ of $\beta$. If $\beta'=\beta$, $\mathcal{B}$ outputs $0$ indicating that $T=e(g,g)^{a^{q+1}s}$. Otherwise $\mathcal{B}$ outputs $1$, which means $T$ is a random elements.}

In the case where $\mu=1$, $\mathcal{A}$ gains no knowledge about $\mu$, therefore, $Pr[\mathcal{B}(\vec{y},T=R)=1|\mu=1]=\frac{1}{2}$. If $\mu=0$, then the ciphertext is valid. We have $$Pr[\mathcal{B}(\vec{y},T=e(g,g)^{a^{q+1}s})=0|\mu=0]=\frac{1}{2}+\epsilon.$$

Therefore, the overall advantage of $\mathcal{B}$ in the $q$-$parallel$ $BDHE$ game is $\frac{\epsilon}{2}$, which is non-negligible.

Since that the decisional $q$-$parallel$ $BDHE$ assumption holds in $G$ and $G_T$, the probability that the adversary breaks the security of our protocol is negligible.

\section{Performance evaluation}\label{sec6}
In this section, we first discuss the numeric results in terms of computation and communication overheads, and then report the implementation results.

\subsection{Numeric Analysis}
\textbf{Computation cost}: In our protocol, AA generates users' private keys according to their attributes. Data are generated and encrypted by the smart object, and decrypted by the users. Data deletion is done by the fog devices. We present the computation cost from the viewpoint of the AA, the fog device, the smart object and the user. For simplicity, $M_G$ and $E_G$ denotes the multiplication and exponentiation in group $G$ and use $M_{G_T}$, $E_{G_T}$ to denote the multiplication and exponentiation in group $G_T$. Denote $P$ the pairing computation. We ignore the computation cost of the hash functions since it is negligible compared with other expensive operations.

  AA is responsible for generating public parameters and private keys for users. Assume a user owns $s$ attributes, the computation cost of AA is $(s+2)E_G+1M_G$. For the smart object, the primary computation is generating the ciphertext of the data key as the cost of the symmetric encryption is little. For a share-generating matrix $M$ with $l$ rows, the dominated cost of the smart object is $3lE_G+lM_G+1P+1M_{G_T}$. The main computation of users is decrypting the encrypted data key. Suppose that the rows in $M$ corresponding to the attributes satisfying the access structure is $i$, the computation cost of the user is $(2P+1M_{G_T}+1E_{G_T})i+2M_{G_T}$.

 Regarding data deletion, the computation cost of the fog device is $2$ exponentiations in $\mathbb{Z}_p$ and $1$ exponentiation in $G$. When validating a deletion proof, the cost of the smart object is $(2P+1M_{G_T}+1E_{G_T})i+2M_{G_T}+1E_G$, where $i$ denotes the number of the rows in $M$ that corresponds to the attributes of the smart object.

\textbf{Communication cost}: After encryption, the smart object uploads the $\{fname,spk,CT,\{D\}_k,(M,\rho)\}$ to the fog device. The size of $\{D\}_k$ is related to the data since data are encrypted with symmetric encryption. The ciphertext size of the data key is $O(l)$, where $l$ is the size of an access formula. During the deletion process, the deletion request contains $4$ group elements in $\mathbb{Z}_p$. The deletion proof returned from the fog device includes two group elements in $\mathbb{Z}_p$. In the verification phase, the smart object needs to retrieve the ciphertext of the data key whose size is $O(l)$, where $l$ denotes the size of the access formula in the ciphertext.

\subsection{Implementation}
In our implementation, we use $SHA$-$1$ as the one-way hash function and $AES$ as the data encryption algorithm. The experiments were conducted with Miracl library \cite{MIRACL} on Intel i5-2450 MQ CPU @ 2.50 GHz. The size of AES keys is $128$ bits and the security parameter is $80$, which satisfies the security requirements.

\begin{figure}
        \centering
                \includegraphics[width=0.5\textwidth]{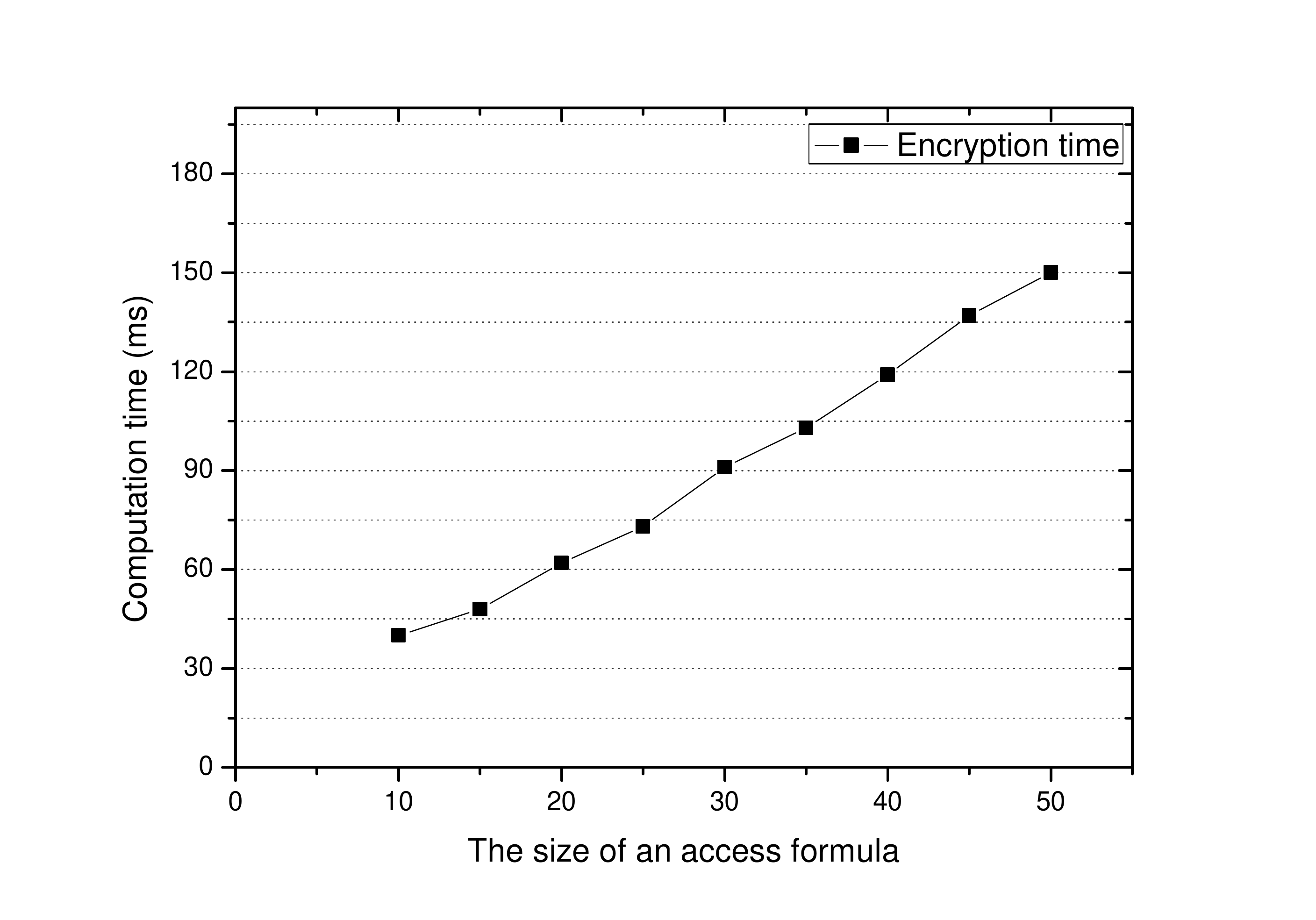}
                \caption{Increasing size the access formula in ciphertext}
                \label{fig:2}
\end{figure}

Since the dominated computation of the smart object is encrypting the data key, we increase the size of the access formula embedded in the ciphertext to assess the encryption cost on the smart object side. In our experiments, we increase the parameters of the share-generating matrix, $l$ and $n$, simultaneously from 10 to 50 with an increment of 10 for each test. We observe that the time cost of the encryption grows as the size of the access formula increases. As shown in Fig. \ref{fig:2}, when there are 50 rows in the $M$, the encryption time is about 150 ms, which is acceptable for the smart object. From the Fig. \ref{fig:3}, we can see that the time cost of the AA for generating a private key for a user is small. When the user owns 10 attributes, the time cost of AA is almost 28 ms. For the user, the decryption time cost increases with the number of attributes of the user. More specifically, the time cost is proportional to the the number of the rows in M, which is determined by the number of user's attributes. The time cost is approximately 180 ms if there are 10 attributes in the user's attribute set.

\begin{figure}
        \centering
                \includegraphics[width=0.5\textwidth]{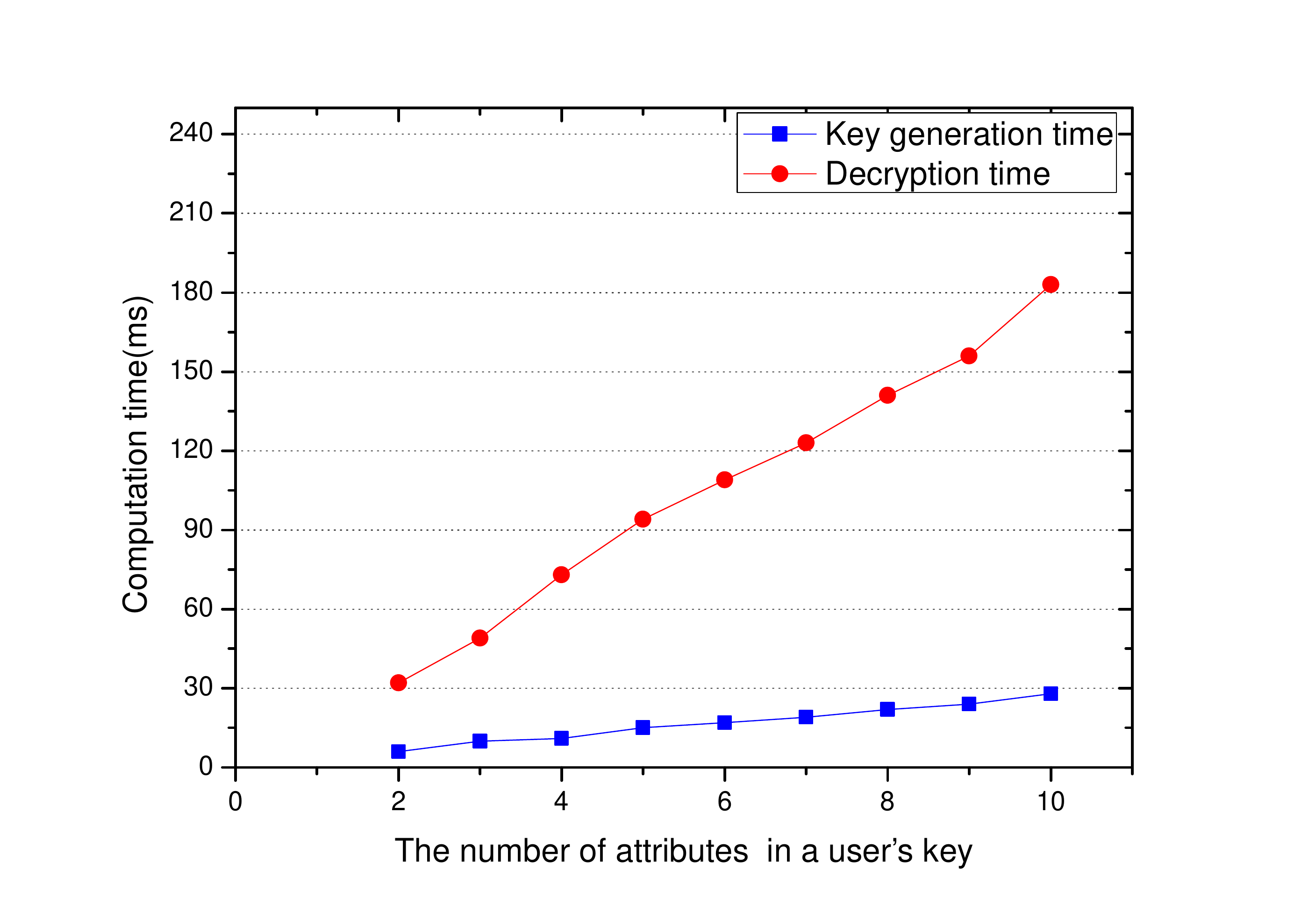}
                \caption{The time of key generation and decryption for increased number of attributes of a user}
                \label{fig:3}
\end{figure}

During the data deletion process, the time to generate the deletion request for the smart object is small since there are only one exponentiation computation. On the fog device side, the time cost of producing a proof is small as well, because only three exponentiation is needed. The main cost of the smart object is to verify the proof returned from the fog device, which is almost the same as the decryption cost. Fortunately, it is a one time computation for a unique file. As shown in Fig. \ref{fig:4}, it consumes about 115 ms, 144 ms, 180 ms respectively if the smart object has 6, 8, 10 attributes, which is bearable to the smart object.

\begin{figure}
        \centering
                \includegraphics[width=0.5\textwidth]{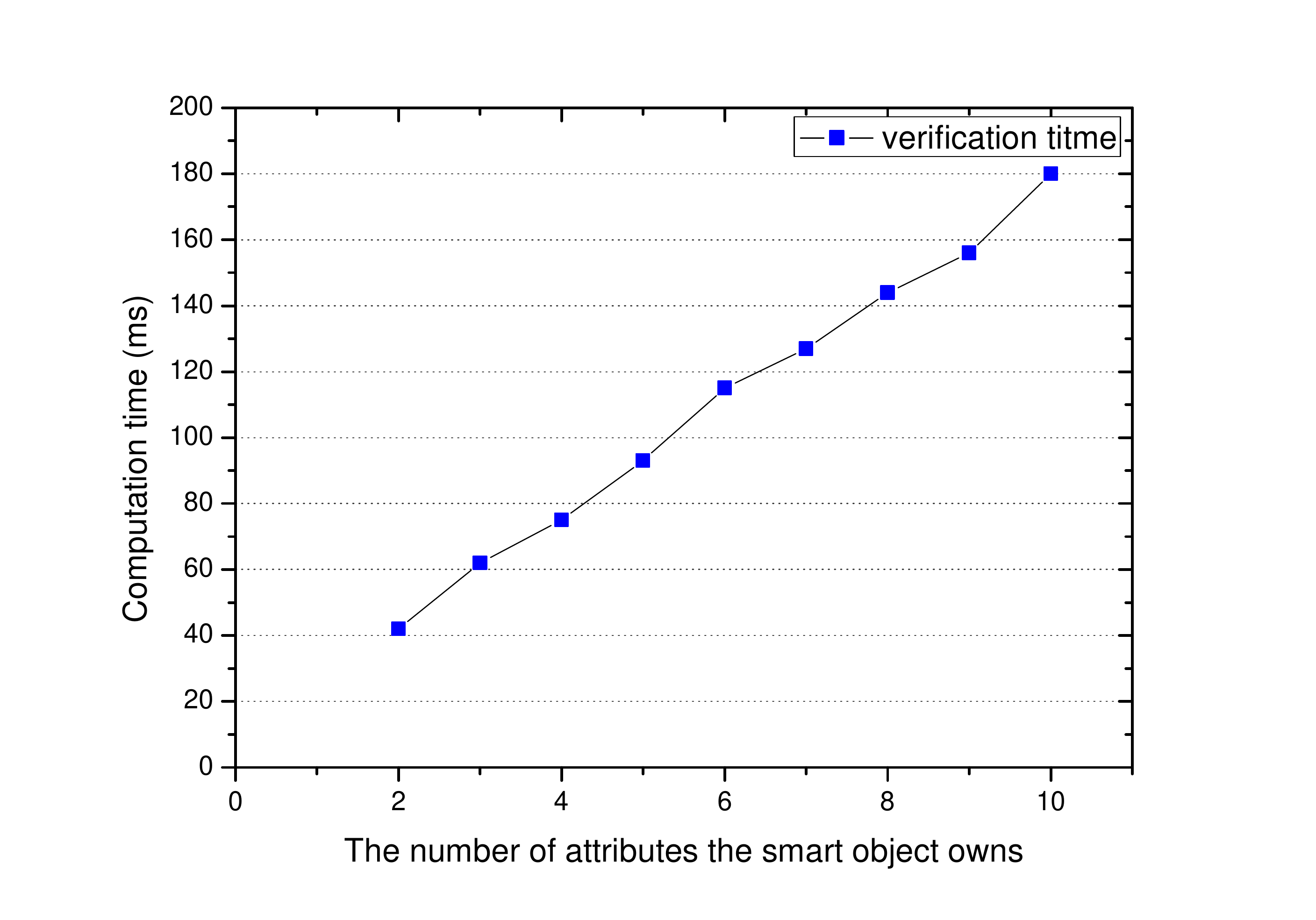}
                \caption{The time of verification for increased number of attributes of a smart object}
                \label{fig:4}
\end{figure}

\section{Conclusion}\label{sec7}
In this paper, we investigated an essential but challenging issue of fine-grained data access and assured data deletion in fog-based industrial applications. To solve the problem, we proposed a new protocol called CPAD, in which smart objects encrypt the data with an access structure and only the users with intended attributes can decrypt it correctly. When the data needs to be deleted, the fog device can do the operations such that none of the users can recover the data key, which makes data irrecoverable. Moreover, CPAD is resilient against unauthorized user colluding attack. The implementations show our protocol can be adopted in real-world applications.



%

%
%
%
%
%

\ifCLASSOPTIONcaptionsoff
  \newpage
\fi

\end{document}